# Large anisotropic magnetoresistance in paramagnetic Fe-doped $Ta_4SiTe_4$

Keita Nagasawa[1,a)], Ryutaro Okuma[1], Jun-ichi Yamaura[1], and Yoshihiko Okamoto[1]

[1]Institute for Solid State Physics, The University of Tokyo, Kashiwanoha 5-1-5, Kashiwa 277-8581, Japan

[a)]Author to whom correspondence should be addressed: nkeita0121@issp.u-tokyo.ac.jp

## Abstract

Large anisotropic magnetoresistance (AMR) usually relies on magnetic ordering, which fixes a preferred moment orientation and couples it to charge transport. Here, we demonstrate that pronounced AMR can instead arise from the anisotropic magnetic-field-induced polarization of dilute local moments in paramagnetic Fe-doped $Ta_4SiTe_4$, a quasi-one-dimensional van der Waals narrow-gap semiconductor. Magnetization measurements reveal Curie-like paramagnetism down to 2 K, with the *c* axis as the magnetic easy axis. The magnetization is described by an effective $S$ = 1/2 model with an anisotropic $g$-factor ratio of $g_c/g_{ab}$ = 1.43(1). With the electrical current applied along the *c* axis, the Fe-doped sample exhibits a large negative magnetoresistance for $B//c$ and only a weak positive response for $B//ab$, whereas the undoped sample shows little magnetoresistance for either field orientation. The resulting AMR ratio reaches –33% at 8 K and 10 T and exhibits a nonmonotonic temperature dependence qualitatively captured by a minimal spin-disorder model. These results show that engineering the local environment of dilute magnetic ions provides a route to large paramagnetic AMR without long-range magnetic order, extending the materials design space for AMR beyond magnetically ordered systems.

Dilute magnetic semiconductors, in which magnetic ions are introduced into nonmagnetic semiconductor hosts, provide an important platform for investigating the coupling between localized magnetic moments and itinerant carriers[1,2]. In prototypical 3*d*-based dilute magnetic semiconductors such as Mn-doped GaAs, carrier-mediated ferromagnetism and associated magnetotransport phenomena have been extensively investigated[3–10]. Among the various magnetotransport phenomena, anisotropic magnetoresistance (AMR), in which magnetic anisotropy is reflected in the electrical resistance, is particularly important because it enables electrical detection of the magnetization direction[11]. Large AMR has been reported in dilute magnetic semiconductors[4,7–10], correlated materials near metal–insulator transitions[12], and antiferromagnetic semiconductors[13]. Across these diverse material systems, large AMR has been observed predominantly in magnetically ordered phases. Although crystalline symmetry permits AMR even in the paramagnetic state, experimental observations of large AMR remain scarce.

Here, we propose a new route to realize large AMR in dilute magnetic semiconductors that does not rely on magnetic ordering: anisotropic magnetic-field-induced polarization of dilute magnetic moments. For an anisotropic local moment, magnetic-field-induced polarization should depend on the direction-dependent effective $g$-factor through the Zeeman splitting. This results in anisotropic suppression of spin-disorder scattering, which is expected to produce large AMR even in the paramagnetic state. Such strongly anisotropic local moments can arise when the orbital degrees of freedom of the magnetic ions are not fully quenched, which can be promoted by a highly anisotropic local coordination environment. However, known examples of strongly anisotropic 3*d* local moments are predominantly found in insulating materials[14–21]. Consequently, the influence of anisotropic dilute 3*d* moments on carrier transport in semiconducting hosts remains largely unexplored.

To explore this possibility, we investigate Fe-doped $Ta_4SiTe_4$, a quasi-one-dimensional narrow-gap semiconductor composed of chains of Si-centered square antiprisms[22–27]. While $Ta_4SiTe_4$ has been intensively studied as a thermoelectric material, the highly anisotropic coordination of the Si site and the existence of isostructural $Ta_4MTe_4$ ($M$ = Cr, Fe, Co, and Ni)[28,29] make it a promising platform for investigating anisotropic magnetotransport induced by dilute $3d$ magnetic dopants. We found that Fe doping produces strongly anisotropic paramagnetism without magnetic ordering down to 2 K and gives rise to large AMR. Remarkably, the observed evolution of the AMR with temperature and magnetic field is qualitatively reproduced by a model of anisotropic magnetic-field-induced polarization based on the effective $g$-factor anisotropy obtained from the magnetization data. These results establish anisotropic magnetic-field-induced polarization of dilute magnetic moments as a route to large AMR without magnetic ordering.

Single crystals of undoped and Fe-doped $Ta_4SiTe_4$ were grown by a vapor-phase method (see Supplemental Note S1)[23]. The crystals were characterized by single-crystal X-ray diffraction (XRD, Rigaku XtaLAB P200, Mo $K\alpha$ radiation). The crystal structure was analyzed using JANA2006[30] and visualized using VESTA[31]. Electrical resistivity measurements were performed using a Physical Property Measurement System (PPMS, Quantum Design). Magnetization measurements were performed using an MPMS-3 (Quantum Design). To minimize sample-to-sample variation, magnetization measurements for different field directions were carried out on the same single crystal by rotating the crystal with respect to the applied magnetic field.

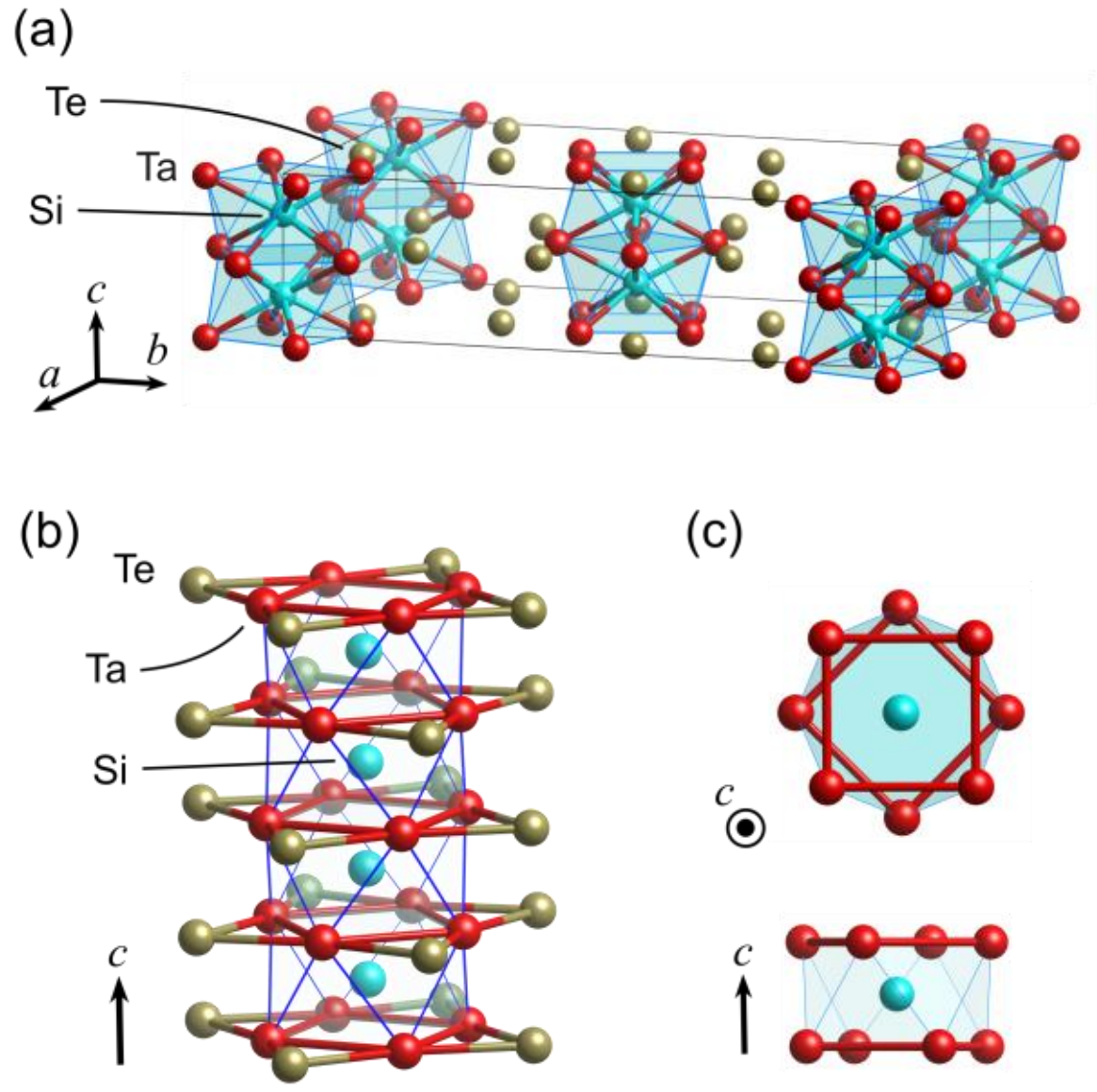

FIG. 1. Structural characterization of undoped $Ta_4SiTe_4$. (a) Crystal structure determined by single-crystal XRD. Red, light blue, and gold spheres represent Ta, Si, and Te atoms, respectively. The space group is *Pbam*, with $a$ = 10.531(3) Å, $b$ = 18.277(5) Å, and $c$ = 4.7997(10) Å. (b) A chain of Si-centered $Ta_8$ square antiprisms. (c) Top (upper) and side (lower) views of the Si-centered $Ta_8$ coordination polyhedron. The local arrangement of the eight Ta atoms around Si closely approximates a square antiprism with $D_{4d}$ symmetry and possesses an approximate noncrystallographic $S_8$ improper-rotation axis.

Needle-like single crystals were obtained for both undoped and Fe-doped $Ta_4SiTe_4$. Single-crystal XRD measurements confirmed that the crystals adopt the reported structure[22] [Fig. 1(a); see Supplemental Note S1]. The structure contains chains of face-sharing, Si-centered square antiprisms extending along the $c$ axis [Fig. 1(b)]. The local coordination environment around the Si site of an undoped crystal is shown in Fig. 1(c). The Si atom is surrounded by eight Ta atoms, forming a nearly ideal square-antiprismatic coordination geometry. Although the dilute Fe concentration prevents definitive identification of the Fe ions by XRD, the magnetic and magnetotransport results presented below are consistent with the presence of Fe ions in $Ta_4SiTe_4$.

The magnetic susceptibility of undoped $Ta_4SiTe_4$ was negative and nearly temperature independent, consistent with diamagnetic behavior, as shown in the inset of Fig. 2(a). A small magnetic anisotropy was observed, $\chi_{ab}^{\mathrm{undoped}} = -6.2(4) \times 10^{-4}$ cm$^3$/mol and $\chi_{c}^{\mathrm{undoped}} = -3.4(5) \times 10^{-4}$ cm$^3$/mol, likely reflecting the quasi-one-dimensional electronic structure. In contrast, Fe-doped $Ta_4SiTe_4$ exhibits Curie-like paramagnetic behavior, in which the magnetic susceptibility increases upon cooling, as shown in Fig. 2(a). No anomaly indicative of magnetic ordering was observed down to 2 K. Despite the absence of magnetic ordering, pronounced magnetic anisotropy develops in the paramagnetic state, with the $c$ axis serving as the magnetic easy axis.

The magnetization curves measured for $B//c$ and $B//ab$ at 2, 3, 4, 5, 8, and 15 K were fitted simultaneously using an effective pseudospin-1/2 Brillouin-function model.

$$M_i(B,T) = \frac{ng_i}{2}\tanh\left(\frac{g_i\mu_B B}{2k_B T}\right) + \chi_{0,i}^{\mathrm{Fe}} B \quad (1)$$

where $i = ab$ or $c$, $n$ is the number of spins per formula unit, $g_i$ is the direction-dependent effective $g$-factor, and $\chi_{0,i}^{\mathrm{Fe}} B$ represents the temperature-independent linear background. A common $n$ was used for both field orientations, whereas $g_i$ and $\chi_{0,i}$ were allowed to be anisotropic. The global fit gives $n = 3.2(2) \times 10^{-3}$ per formula unit, $g_{ab} = 2.8(1)$, and $g_c = 4.0(1)$, corresponding to $g_c/g_{ab} = 1.43(1)$. The fitted backgrounds, $\chi_{0,ab}^{\mathrm{Fe}} = -6.0(1) \times 10^{-4}$ cm$^3$/mol and $\chi_{0,c}^{\mathrm{Fe}} = -3.6(3) \times 10^{-4}$ cm$^3$/mol, are close to the susceptibilities of undoped $Ta_4SiTe_4$. The nominal starting composition does not represent the actual Fe content. The fitted $n$ therefore denotes an effective concentration of magnetically active moments rather than a direct chemical determination of the Fe concentration. Although the fitted values of $g_i$ and $n$ depend on the assumed value of $S$, the anisotropy ratio $g_c/g_{ab}$ remains nearly independent of this choice (see Supplemental Note S2).

From Eq. (1), the data obtained at different temperatures approximately collapse onto a common curve for each field orientation when plotted as $M_i - \chi_{0,i}^{\mathrm{Fe}} B$ versus $B/T$, as shown in Fig. 2(b). The solid curves represent Eq. (1) using the global-fit parameters. The separation between the two orientation-dependent scaling curves originates from the anisotropic effective $g$-factors and demonstrates that the dilute moments are polarized more efficiently for $B//c$. Using these parameters, $M_i/B$ at $B$ = 1 T was calculated from Eq. (1) and compared with the low-temperature susceptibility in Fig. 2(a). The calculated curves capture the overall Curie-like increase and its magnetic anisotropy. These results are consistent with the response of dilute, isolated, anisotropic local moments described by a minimal $S$ = 1/2 model.

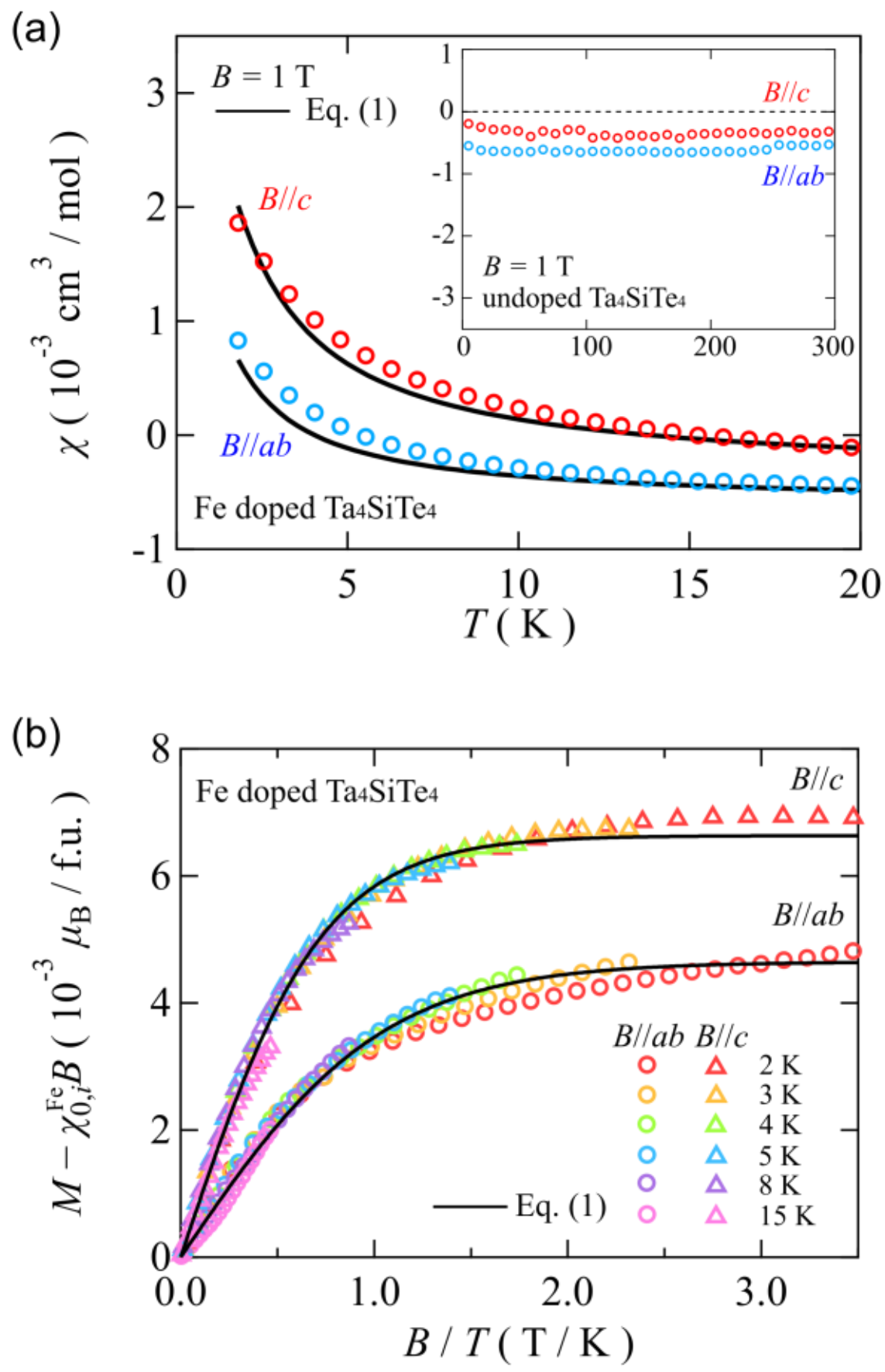

FIG. 2. Magnetic properties of undoped and Fe-doped $Ta_4SiTe_4$. (a) Magnetic susceptibility of Fe-doped $Ta_4SiTe_4$ measured at $B = 1$ T for *B*//*ab* and *B*//*c*; the inset shows the corresponding data for undoped $Ta_4SiTe_4$. The solid curves for the Fe-doped sample were calculated from the global fits to $M_i(B, T)$. (b) Background-subtracted magnetization versus *B*/*T* at 2–15 K. The solid curves are global fits to Eq. (1).

Figure 3(a) shows the temperature dependence of the electrical resistivity measured with *I*//*c*. Undoped $Ta_4SiTe_4$ exhibits a temperature dependence consistent with that reported previously[23]. In contrast, the resistivity of the Fe-doped sample increases more rapidly upon cooling, suggesting that Fe doping enhances carrier scattering at low temperatures. The influence of Fe doping is even more pronounced in the magnetoresistance.

To investigate the anisotropic magnetotransport, we measured the angular dependence of the resistivity by rotating the magnetic field within a plane containing the *c* axis, while keeping the electric current along the chain direction (*I*//*c*), as illustrated in the inset of Fig. 3(a). Here, $\theta$ is defined as the angle between the magnetic field and the *c* axis. Figure 3(b) shows the magnetic-field dependence of the resistivity of Fe-doped $Ta_4SiTe_4$ for the two limiting field orientations. A small positive magnetoresistance was observed for $\theta = 90^\circ$ ($B//ab$), whereas a large negative magnetoresistance was observed for $\theta = 0^\circ$ ($B//c$). We define the angle-dependent AMR as $\mathrm{AMR}(\theta) =$ $\left(\rho(\theta) - \rho(90°)\right)/\left(\left(\rho(0°) + \rho(90°)\right)/2\right)$. We further define the AMR ratio $r_{\mathrm{AMR}}$ as $r_{\mathrm{AMR}} = \mathrm{AMR}(0°)$. $\mathrm{AMR}(\theta)$ and $r_{\mathrm{AMR}}$ are expressed as a percentage. Figure 3(c) shows $\mathrm{AMR}(\theta)$ for undoped and Fe-doped $Ta_4SiTe_4$. The undoped sample exhibits only a weak angular dependence, whereas the Fe-doped sample exhibits a large negative AMR at $\theta = 0°$, corresponding to $B//c$. At 8 K and 10 T, $r_{\mathrm{AMR}}$ reaches –33%. In the present measurement geometry, the current is always applied along the $c$ axis. Thus, $B//c$ and $B//ab$ correspond to $B//I$ and $B \perp I$, respectively. For conventional orbital magnetoresistance arising from the Lorentz force, a larger *positive* magnetoresistance is expected for $B \perp I$, whereas this contribution should be suppressed for $B//I$. The large *negative* magnetoresistance observed in Fe-doped $Ta_4SiTe_4$ for $B//I$ points to an anisotropic contribution to AMR that cannot be explained by a conventional Lorentz-force contribution alone. Thus, Fe-doped $Ta_4SiTe_4$ exhibits a strongly field-direction-dependent magnetoresistance while remaining paramagnetic.

To clarify the origin of the observed large negative AMR in Fe-doped $Ta_4SiTe_4$, we examine whether the dependence of the resistivity on temperature and magnetic field can be understood in terms of spin-disorder scattering. In the simplest model[32,33], polarization reduces the spin-disorder resistivity according to $\Delta\rho^{\mathrm{sd}}(T, B) = f\rho_0 P^2(T, B)$, with $f < 0$, where $\rho_0(T)$ is the measured zero-field resistivity and $P(T, B)$ is the normalized spin polarization given by the magnetization divided by the saturation magnetization. In the present case, the Fe moments are more readily polarized for $B//c$ because of the easy-axis anisotropy, leading to a stronger suppression of spin-disorder scattering. Including this term, the resistivity can be written as

$$\rho_i(T, B) = \rho_0(T) + \Delta\rho_i^{\mathrm{sd}}(T, B) = \rho_0(T)\left(1 + f\tanh^2\left(\frac{g_i \mu_B B}{2k_B T}\right)\right) \quad (2)$$

Here, $f$ is a common coefficient independent of temperature and field direction. The polarization $P(T, \boldsymbol{B})$ was calculated from the Brillouin-function component of Eq. (1), excluding the linear background, using the fitted $g$-factors $g_{ab} = 2.8$ and $g_c = 4.0$. The AMR ratio, $r_{\mathrm{AMR}}(T, B)$, is then calculated from the modeled resistivities for the two field orientations as

$$r_{\mathrm{AMR}} = \frac{f(P_c^2 - P_{ab}^2)}{1 + \frac{f}{2}(P_{ab}^2 + P_c^2)} \quad (3)$$

Figure 3(d) compares the temperature dependence of $r_{\mathrm{AMR}}$ at 10 T obtained from the measured resistivities with that calculated using Eq. (3). A distinctive feature of the experimental data is the nonmonotonic temperature dependence: the magnitude of $r_{\mathrm{AMR}}$ increases upon cooling, reaches a pronounced maximum near 8 K, and decreases again at lower temperatures. Despite its simplicity, the model captures the unusual peak behavior, supporting anisotropic spin-disorder scattering as the origin of this behavior. At 10 T, the model predicts a minimum in $r_{\mathrm{AMR}}$ at approximately 8.5 K, close to the experimentally observed minimum near 8 K. Substantial deviations remain at both the lowest and highest temperatures, indicating additional transport contributions beyond the present minimal model (see Supplemental Note S3).

The nonmonotonic AMR follows naturally from the field and temperature dependence of the Fe-moment polarization. In the present model (Eq. (3)), both $P_c$ and $P_{ab}$ are small at high temperature, resulting in $r_{\text{AMR}}$ close to zero. Upon cooling, $P_c$ increases more rapidly than $P_{ab}$ because $g_c > g_{ab}$, thereby enhancing $|r_{\text{AMR}}|$ and producing an increasingly negative AMR. At still lower temperatures, however, the moments approach saturation for both field directions, causing $P_c^2 - P_{ab}^2$ to decrease again. The magnetic contribution to the AMR therefore reaches a maximum at an intermediate temperature.

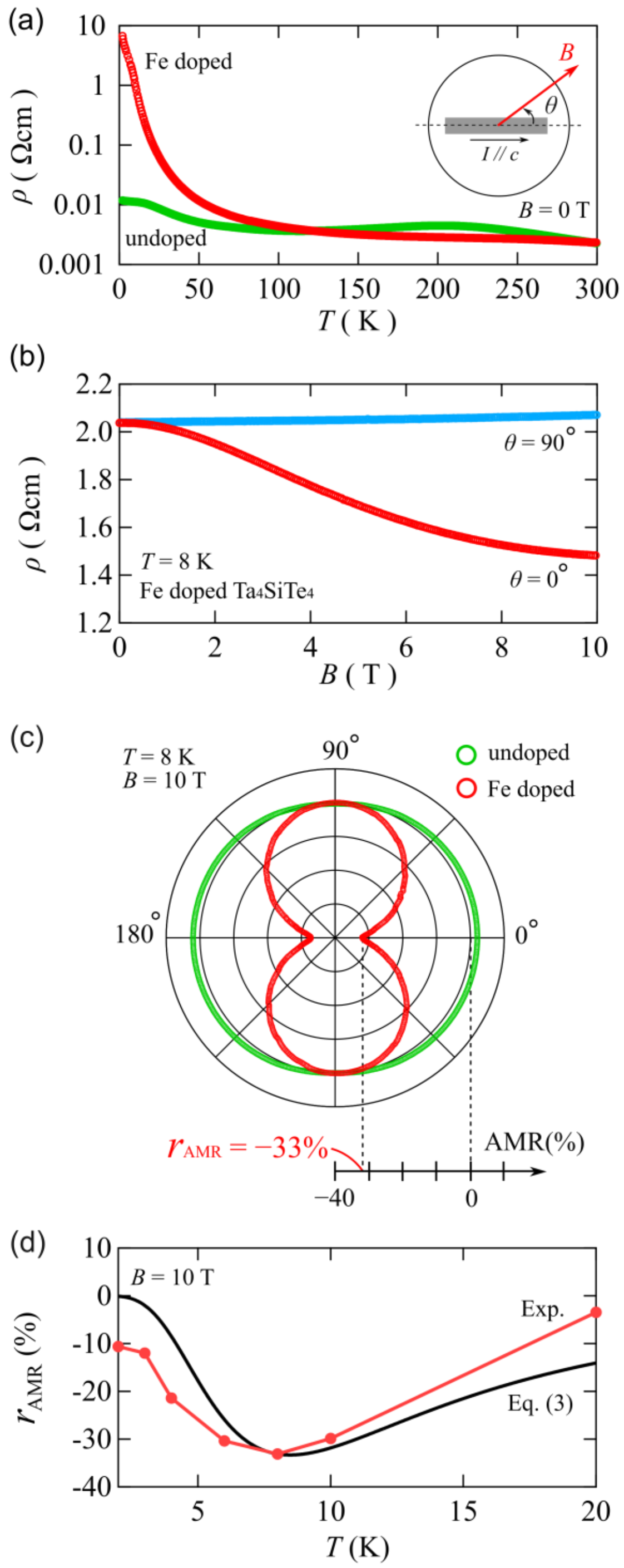

FIG. 3. Transport properties of undoped and Fe-doped $Ta_4SiTe_4$. (a) Temperature dependence of the electrical resistivity of undoped (green) and Fe-doped (red) $Ta_4SiTe_4$ measured with *I*//*c*. Inset: Measurement geometry for the angle-dependent magnetoresistance. (b) Magnetic-field dependence of the resistivity of Fe-doped $Ta_4SiTe_4$ measured at $T = 8$ K for $\theta = 90°$ (blue) and $\theta = 0°$ (red). (c) AMR($\theta$) of undoped (green) and Fe-doped (red) $Ta_4SiTe_4$ measured at $T = 8$ K and $B = 10$ T. (d) Temperature dependence of $r_{\mathrm{AMR}}$ at $B = 10$ T obtained experimentally (red symbols) and calculated using Eq. (3) (black solid line). The parameter $f = -0.745$ was determined by least-squares fitting to the experimental data.

Finally, to place the present results in context, we compare the observed AMR magnitude with reported values for representative magnetic materials. Although AMR values reported using different normalization conventions are not directly comparable, the value of $|r_{AMR}| = 33\%$ observed here substantially exceeds the approximately 0.2−2% AMR reported for typical elemental $3d$ ferromagnets[34], the AMR of several percent reported for ferromagnetic (Ga,Mn)As[7–10], and the 1.5% paramagnetic AMR of $SrRuO_3$[35]. Although larger values have been reported for an optimized ferromagnetic $Ni_{80}Co_{20}$ alloy (~68%)[36] and antiferromagnetic $EuTe_2$ ($\sim 4 \times 10^4\%$)[13], these responses rely on magnetically ordered states; in $EuTe_2$, the colossal AMR is associated with a field-direction-driven band reconstruction and metal–insulator transition. By contrast, the present results support a mechanism based on the direction-dependent suppression of spin scattering through the anisotropic Zeeman polarization of dilute local moments, without invoking long-range magnetic order or a field-induced phase transition. The anisotropic $g$-factors of these moments are consistent with the combined effects of spin–orbit coupling and a strongly uniaxial local crystal-field environment. These findings suggest a possible route to large paramagnetic AMR in dilute magnetic semiconductors: combining anisotropic local-moment polarization with a substantial contribution of spin scattering to the resistivity. Semiconductor hosts that provide strongly anisotropic coordination environments for dilute magnetic ions may therefore offer promising candidates for exploring this mechanism.

In summary, we investigated the magnetic and magnetotransport properties of Fe-doped $Ta_4SiTe_4$, a quasi-one-dimensional van der Waals semiconductor containing dilute Fe dopants. The Fe-doped sample exhibits anisotropic Curie-like paramagnetism with the crystallographic $c$ axis as the magnetic easy axis, together with anisotropic effective $g$-factors characterized by $g_c \,/\, g_{ab} = 1.43(1)$. Furthermore, despite the absence of magnetic ordering down to 2 K, a large negative AMR of −33% was observed in the paramagnetic state. The characteristic nonmonotonic temperature dependence of $r_{\mathrm{AMR}}$ is qualitatively reproduced by a minimal spin-disorder-scattering model incorporating the anisotropic field-induced polarization of the dilute Fe moments. These findings suggest that employing strongly uniaxial yet locally high-symmetry coordination environments to retain an appreciable orbital contribution to the moments of dilute $3d$ ions may provide a promising strategy for realizing strongly anisotropic magnetotransport without long-range magnetic order.

We are grateful to K. Moriyama, K. Kojima, H. Isshiki, M. Hara, and M. Hashisaka for helpful discussions. This work was supported by JSPS KAKENHI (Grant Nos. 23K26524, 26H02231, and 26H00640), JST ASPIRE (Grant No. JPMJAP2314), JST SPRING (Grant No. JPMJSP2108), and JST ARiSE (Grant No. JPMJAR2628).

## AUTHOR DECLARATIONS

### Conflict of Interest

The authors have no conflicts to disclose.

### Author Contributions

K. Nagasawa: Data curation; Formal analysis (equal); Investigation (lead); Visualization; Writing – original draft; Writing – review & editing (equal).

R. Okuma: Conceptualization; Formal analysis (equal); Investigation (supporting); Supervision (equal); Resources (equal); Writing – review & editing (equal); Funding acquisition (equal).

J. Yamaura: Resources (equal); Writing – review & editing (equal).

Y. Okamoto: Supervision (equal); Resources (equal); Writing – review & editing (equal); Funding acquisition (equal).

## DATA AVAILABILITY

The data that support the findings of this study are available from the corresponding author upon reasonable request.

## SUPPLEMENTARY NOTES

### S1. Synthesis and characterization of undoped and Fe-doped $Ta_4SiTe_4$

Single crystals of $Ta_4SiTe_4$ and Fe-doped $Ta_4SiTe_4$ were grown by a vapor phase method. For the parent $Ta_4SiTe_4$, Ta, Si, and Te powders were mixed in a nominal molar ratio of 2:1:2, and $TeCl_4$ was added as a transport agent. The starting materials were loaded into a quartz ampoule, which was evacuated to approximately $10^{-3}$ Pa. Before the ampoule was sealed, it was flame-heated for approximately 10 min to remove residual moisture from its inner surface. The sealed ampoule was first held at 600 °C for 24 h, then held at 1050 °C for 96 h, and finally furnace-cooled to room temperature. For the Fe-doped samples, a nominal starting molar ratio of Ta:Si:Fe:Te = 2:0.7:0.3:2 was used. Crystallographic data and fractional atomic coordinates are summarized in Tables S1-1 and S1-2, respectively.

**Table S1-1**. Crystallographic data and structure-refinement parameters for $Ta_4SiTe_4$ obtained from single-crystal X-ray diffraction.

| | |
|---|---|
| Formula | $Ta_4SiTe_4$ |
| Crystal system | Orthorhombic |
| Space group | *Pbam* |
| $a$ (Å) | 10.531(3) |
| $b$ (Å) | 18.277(5) |
| $c$ (Å) | 4.7997(10) |
| $V$ (Å$^3$) | 923.8(4) |
| Radiation | Mo $K\alpha$ (λ = 0.71075 Å) |
| Temperature (K) | 293 |
| $\mu$ (mm$^{-1}$) | 59.644 |
| $\theta_{\max}$ (deg.) | 36.78 |
| Index ranges | $-17 \leq h \leq 17, -30 \leq k \leq 30, -7 \leq l \leq 7$ |
| Independent reflections | 2470 |
| Reflections with $I > 3\sigma(I)$ | 1402 |
| Parameters refined | 54 |
| $F_{000}$ | 2056 |
| $R_{\text{obs}}$ ($I > 3\sigma$) | 0.0571 |
| $wR_{\text{obs}}$ ($I > 3\sigma$) | 0.0470 |
| Goodness-of-fit | 1.60 |
| Crystal size ($\mu$m) | $20 \times 20 \times 20$ |
| Largest peak/hole (eÅ$^{-3}$ ) | +3.55/−4.96 |

**Table S1-2**. Fractional atomic coordinates and equivalent displacement parameters. $U_{\mathrm{eq}}$ is defined as one third of the trace of the orthogonalized $U_{ij}$ tensor.

| Atom | $x$ | $y$ | $z$ | $U_{\mathrm{eq}}$(Å$^2$) |
|---|---|---|---|---|
| Ta1 | 0.06575(10) | 0.11952(6) | 0.5000 | 0.0113(3) |
| Ta2 | 0.20862(10) | 0.96310(6) | 0.5000 | 0.0115(3) |
| Ta3 | 0.19337(10) | 0.05853(6) | 0 | 0.0114(3) |
| Ta4 | 0.89886(10) | 0.11080(6) | 0 | 0.0115(3) |
| Te1 | 0.35850(16) | 0.93480(10) | 0 | 0.0150(5) |
| Te2 | 0.82758(17) | 0.19049(10) | 0.5000 | 0.0150(5) |
| Te3 | 0.33437(16) | 0.10057(10) | 0.5000 | 0.0154(5) |
| Te4 | 0.11040(17) | 0.20634(9) | 0 | 0.0151(5) |
| Si1 | 0 | 0 | 0.2492(16) | 0.0129(19) |

**S2. Results of global fits to the magnetization**

Although larger assumed $S$ values yield modestly smaller RMS residuals, the available data do not uniquely constrain $S$. We therefore use $S$ = 1/2 only as a minimal effective two-level parameterization. This choice should be regarded as an effective parameterization rather than a microscopic identification of the Fe spin state. Importantly, both the anisotropy ratio $g_c/g_{ab}$ and the temperature-independent susceptibilities $\chi_{0,i}$ remain nearly unchanged over the examined range of $S$, demonstrating that the principal conclusions are insensitive to this assumption.

**Table S2-1.** Dependence of the fitted magnetic parameters on the assumed $S$. The $M(B,T)$ curves measured at $T$ = 2, 3, 4, 5, 8, and 15 K were fitted simultaneously with $S$ fixed at each value using the Brillouin function. The moment density $n$ was constrained to be common to the two field orientations, whereas $g_{ab}$, $g_c$, $\chi_{0,ab}$, and $\chi_{0,c}$ were independently optimized.

| $S$ | $n$ ($10^{-3}$/ f. u. ) | $g_{ab}$ | $g_c$ | $g_c/g_{ab}$ | $\chi_{0,ab}$ ($10^{-4}$ cm$^3$/mol) | $\chi_{0,c}$ ($10^{-4}$ cm$^3$/mol) | RMS residual ($10^{-4}\mu_B$/f. u. ) |
|---|---|---|---|---|---|---|---|
| 1/2 | 3.2(2) | 2.8(1) | 4.0(1) | 1.43(1) | −6.0(1) | −3.6(3) | 1.467 |
| 1 | 2.1(1) | 2.15(9) | 3.0(1) | 1.43(1) | −6.0(1) | −3.7(3) | 1.350 |
| 3/2 | 1.8(1) | 1.72(6) | 2.4(1) | 1.43(1) | −6.0(2) | −3.7(3) | 1.284 |
| 2 | 1.7(1) | 1.43(5) | 2.05(8) | 1.43(1) | −6.0(2) | −3.7(3) | 1.251 |
| 5/2 | 1.6(1) | 1.22(4) | 1.75(7) | 1.43(1) | −6.0(2) | −3.7(3) | 1.235 |

**S3. Results of global fits to the field-dependent resistivity**

We globally fitted the field-dependent resistivity measured at $T$ = 2, 3, 4, 6, 8, 10, and 20 K over $0 \leq B \leq 10$ T using

$$\rho_i(T,B) = \rho_0 + \Delta\rho_i^{\mathrm{sd}} + \rho_i^{\mathrm{orb}} = \rho_0(T)\left[1 + \alpha \tanh^2\left(\frac{g_i \mu_B B}{2k_B T}\right) + \beta_i(T) B^2\right] \quad \text{(S3.1)}$$

Here, $\rho_0(T)$ is the measured zero-field resistivity, $\alpha < 0$ is a common coefficient independent of temperature and field direction, and $\beta_i(T) \geq 0$ is a temperature- and direction-dependent coefficient describing positive orbital magnetoresistance phenomenologically. The polarization $P(T, \boldsymbol{B})$ was calculated from the Brillouin-function component of Eq. (1), excluding the linear background, using the fitted $g$ factors $g_{ab} = 2.8$ and $g_c = 4.0$. For each dataset, $\rho_0(T)$ was fixed to the measured resistivity at $B$ = 0 T. Thus, the global fit contained 14 quadratic coefficients and one common magnetic-scattering coefficient. The polarization was evaluated using the measured temperature at each data point. The fit yielded $\alpha = -0.180$. The negative sign corresponds to a reduction in resistivity with increasing spin polarization. The fitted $\beta_{ab}$ and $\beta_c$ values are shown in Figs. S3-1(a)–S3-1(g).

The model qualitatively captures key features of the observed AMR, including its large negative magnitude and pronounced enhancement at intermediate temperatures. The model is consistent with a contribution from anisotropic field-induced polarization, although the freely varying orbital terms and systematic residuals preclude a unique quantitative assignment. Nevertheless, systematic deviations remain in the individual resistivity curves, particularly for *B*//*ab* and at the lowest and highest temperatures investigated. These discrepancies highlight the limitations of the simplified transport model, which assumes a common magnetic-scattering coefficient and a purely quadratic orbital contribution.

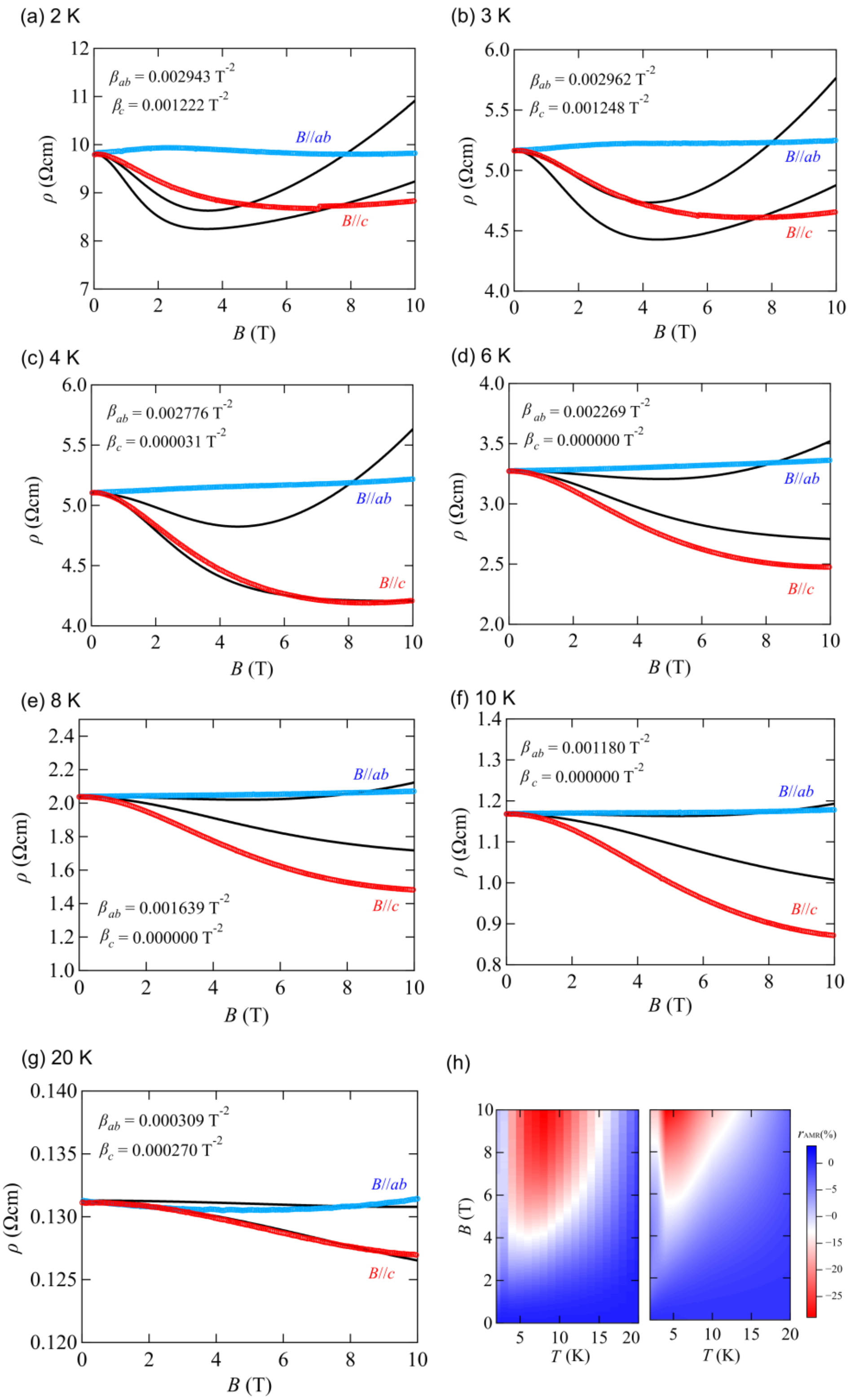

**Fig. S3-1.** (a)−(g) Magnetic-field dependence of the resistivity of Fe-doped $Ta_4SiTe_4$ at 2, 3, 4, 6, 8, 10, and 20 K, measured with *I*//*c*. Blue open circles and red open circles represent the experimental data for *B*//*ab* and *B*//*c*, respectively. Black solid curves show the global fits using Eq. (S3.1). (h) Contour maps of $r_{\mathrm{AMR}}(T,\boldsymbol{B})$ obtained from the measured resistivities (left) and calculated using Eq. (S3.1) (right).